\documentclass[12pt,preprint]{aastex}

\usepackage{emulateapj5}
\usepackage{apjfonts}

\newcommand{\algol}{\mbox{$\beta$~Per}}

\slugcomment{To appear in ApJ, ???, ???}

\shorttitle{Evidence for superhumps in Algol}
\shortauthors{Retter, Richards \& Wu}

\begin{document}

\title {Evidence for Superhumps in the Radio Light Curve of Algol and 
a New Model for Magnetic Activity in Algol Systems}



\author{Alon Retter$^{1,2}$, Mercedes T. Richards$^{1}$, Kinwah Wu$^{3}$}

\vskip 0.4 cm

\affil{$^{1}$Department of Astronomy and Astrophysics, Pennsylvania State 
University, 525 Davey Lab, University Park, PA \\
16802-6305; retter@astro.psu.edu, mrichards@astro.psu.edu
\vskip 0.1 cm
$^{2}$School of Physics, University of Sydney, NSW 2006, Australia
\vskip 0.1 cm
$^{3}$Mullard Space Science Laboratory, University College London,
Holmbury St Mary, Surrey RH5 6NT, UK; kw@mssl.ucl.ac.uk}
\vskip 0.1 cm

\begin{abstract}
Extensive radio data of two Algol systems and two RS~CVn binaries were 
re-analyzed. We found evidence for a new periodicity that we interpret 
as a superhump in $\algol$=Algol, in which it may have been expected 
according to its semi-detached nature and low binary mass ratio. The 
concluded presence of an accretion disk (or an annulus) is consistent 
with previous studies of optical data and numerical simulations of 
$\algol$. In our model, the 50-d period, previously found in the radio 
data of $\algol$, is explained as the apsidal precession of the elliptical 
accretion disk / annulus. If our interpretation is correct, this is the 
first detection of the superhump phenomenon in the radio and the first 
observation of superhumps in Algol systems. According to our result, 
the accretion disk / annulus in $\algol$ precesses in spite of its 
non-Keplerian nature and therefore this phenomenon is not restricted to 
the classical Keplerian accretion disks in compact binaries.


We propose that in Algol systems with short orbital periods, which 
have accretion disks / annuli, the disk is magnetically 
active as well as the cool secondary star. The magnetic field in the 
disk originates from amplification of the seed field in the magnetized 
material transferred from the secondary. The disk and stellar fields 
interact with each other, with reconnection of the field lines causing 
flares and particle acceleration. Relativistic particles are trapped 
in the field and directed toward the polar regions of the secondary 
star because of the dipole structure of its magnetic field. These 
systems are, therefore, somewhat analogous to the RS~CVn systems, 
which have two magnetically interacting stars. Our proposed model for 
the magnetic activity in Algol systems provides a simple explanation 
to the observed properties of $\algol$ in the radio wavelengths, and 
in particular, to the presence of quiescent gyrosynchrotron emission 
near the polar region of the secondary star, where electrons are 
difficult to be confined if the field lines are open as in normal 
single magnetic stars. It further explains the recent discovery that 
the Doppler shifts of the emission lines in the X-ray data of $\algol$ 
are somewhat lower than what is expected from the orbital motion of 
the secondary star. We propose that the superhump variation in the 
radio is generated by enhanced reconnection when the elongated side 
of the elliptic accretion disk is the closest to the cool star. This 
leads to flares and enhancement in particle acceleration and is 
manifested as stronger gyrosynchrotron radiation. 

The observed superhump period, at 3.037$\pm$0.013 d, $\sim$6\% longer 
than the orbital period, was used to deduce a binary mass ratio of 
$\sim$0.23 in $\algol$, which is consistent with previous studies.
Our finding opens the possibility to extend the superhump phenomenon 
to Algol systems and to test the theories of precessing accretion disks 
in various types of interacting binaries. It provides a new method to 
estimate the mass ratios in these binaries. It also offers new insights 
and improves our understanding of the complicated magnetic interaction 
and feedback between mass transfer dynamics, time-dependent disk 
accretion, and feedback and induced magnetic activity in the Algols 
and related systems. 
\end{abstract}



\section{Introduction}

In semi-detached systems, one star fills its Roche-Lobe and transfers 
mass to the other. There are several different groups of interacting 
binaries that are classified according to the properties of their 
component stars. In cataclysmic variables (CVs), a white dwarf accretes 
mass from a red dwarf \citep{W1995}. In low mass X-ray binaries (LMXBs), 
the mass-receiver star is a black hole or a neutron star and the donor 
is typically a low mass solar-like star \citep{LPH1995}. Algol systems 
contain a hot main-sequence star and a cool subgiant or giant companion 
\citep[e.g.,][]{PP1984}. The mass transfer in interacting binaries is 
usually sustained through an accretion disk which is formed around the 
mass-receiver star. In CVs and LMXBs very often the disk is the dominant 
light source in the optical regime. 


Many CVs show in their optical light curves quasi-periodicities a 
few percent different than the orbital periods \citep{W1995, P1999, 
RN2000, PTK2003}. For historical reasons, these periods are known as
superhumps. These humps were initially observed in the bright 
outbursts (superoutbursts) of SU~UMa systems, a subclass of dwarf 
novae, which are a subgroup of CVs. \citet{OC1996} tested and 
confirmed the optical detections of superhumps in three LMXBs. 
\citet{HKM2001} suggested that many more LMXBs share this property. 
There is, however, only one clear detection (in the LMXB 
V1405~Aql=X1916-053) of superhumps in the X-ray regime \citep{RCB2002}. 
When the superhump periods are longer than the orbital periods they 
are termed `positive superhumps'. Similarly, periods shorter than the 
orbital periods are called `negative superhumps'. Hereafter, we refer 
to superhumps as positive superhumps, which are the more common phenomenon. 
The superhump is understood as the beat periodicity between the 
orbital period and the apsidal precession of an elliptical accretion 
disk \citep{O1985, W1988}. It is argued that superhumps can only occur 
in systems with small mass ratios: M$_{donor}$/M$_{receiver}\leq$ 0.33 
\citep{WK1991, M2000}, although \citet{RHA2003} raised some questions 
about this limit by finding superhumps in TV~Col, a CV with a relatively 
large orbital period, which probably implies a mass ratio above the 
theoretical limit. 


Algol systems are named after the proto-type Algol ($\algol$) and may 
be progenitors of CVs \citep{IT1985}. Algol binaries typically contain
a hot B--A main-sequence star and a cool F--K subgiant or giant which 
show a substantial magnetic activity \citep[e.g.,][]{PP1984, RWG2003}. 
As in CVs and LMXBs, the mass transfer in the Algol systems with orbital 
periods longer than about 5 days is sustained through an accretion disk. 
In systems with shorter periods the disk may be transient, and in some 
cases it does not follow the Keplerian velocity field (v $\propto 
r^{-1/2}$) as in compact binaries, so it was termed `an annulus' 
\citep{PP1984, R1992, R1993, RA1999}. Hydrodynamic simulations demonstrate 
that the mass transfer in the Algol binaries can indeed result in the 
formation of an accretion disk, a transient disk or an annulus around 
the mass-receiver star \citep{BRM1995, RR1998} and Doppler tomography 
supports this suggestion \citep{RAB1995, RJS1996}. We note that the 
numerical simulations show that the accretion material around a 
$\algol$-type system can develop an asymmetrical structure that looks 
elliptical in shape \citep[e.g., see fig.~1 of][]{BRM1995}.

Eccentric accretion disks may, therefore, be expected in Algol systems 
whose mass ratios are below the critical value as well. However, in 
these objects the contribution of the disk in the optical band is very 
weak compared with the bright binary stars \citep{R1992}. Thus, to look 
for superhumps it may be preferred to study other wavebands where the 
disk may have a larger impact on the light curve. The radio regime may, 
therefore, offers an opportunity to search for superhumps in Algol 
binaries.

 
In this work, extensive radio observations presented by \citet{RWG2003} 
were re-analyzed. The data cover four binary systems, two of which are 
Algol systems, and the other two belong to the RS~CVn class. The RS~CVn
systems consist of two detached magnetically active stars, and thus 
should not have an accretion disk nor superhumps. The mass ratios of 
the Algol systems studied, $\algol$ and $\delta$~Lib, are 0.22 and 0.35 
respectively \citep{RWG2003}. Therefore, $\algol$ has a mass ratio well 
below the theoretical limit of 0.33 and it should be the only system 
among the four that may show superhumps, although the mass ratio of 
$\delta$~Lib is just above the critical limit. 



$\algol$ (Algol, HD 19356) at V=2.1-3.4 and d=28.76 pc is the brightest 
eclipsing system and the brightest and closest semi-detached binary 
\citep{L1979, MM1998, RA1999, RWG2003}. It has strong emission across 
the wavelength spectrum including the X-ray, ultraviolet, optical, 
infrared and radio \citep[e.g.,][]{LMP1988, R1990, SLS1995, NSB2002, 
RWG2003, SNF2003}. $\algol$ is actually a hierarchical triple system with 
a B8 V main-sequence star, a K2 IV subgiant companion and a binary orbital 
period of 2.8673 days. The tertiary, which is an F1 IV star, orbits around 
the binary system with a period of 680.08 days \citep{HBH1971, RMB1988, 
R1992, R1993, RWG2003}. The binary mass ratio of $\algol$ was estimated 
as 0.217$\pm0.005$ \citep{HBH1971, TL1978} and 0.22$\pm0.03$ 
\citep{RMB1988}. These values locate the system well inside the permitted 
range for superhumps, which would be predicted to be about 4--8\% longer 
than the orbital period \citep{O1985, W1994, M2000}. In this work, we 
present evidence for such a signal.


\section{Analysis}

The radio observations were made between 1995 January and 2000 October 
using the NRAO-Green Bank Interferometer (GBI). The data were collected 
simultaneously in two bands (2.25 and 8.3 GHz). Two Algol systems 
($\algol$ and $\delta$~Lib) and two RS~CVn systems (V711~Tau and UX~Ari)
were observed. The initial analysis was done by \citet{RWG2003}. A careful 
inspection of the power spectrum of $\algol$ (their fig. 29b) suggests 
that this system may have a photometric periodicity several percent 
longer than the orbital period. In addition, the power spectrum of 
$\algol$ is very different from those of the other three objects that 
typically show a strong signal at the orbital period. The new peak in 
the power spectrum of $\algol$ would be naturally interpreted as a 
superhump period. In this work, we highlight this finding, check the 
significance of this signal and show that it is indeed real.


The radio observations of $\algol$ contain 7442 measurements in each 
of the two frequency bands. This object is a strong radio source with 
frequent flares that reach above 1 Jy in the 8.3-GHz band \citep{RWG2003}. 
Figure 1 displays the power spectra \citep{S1982} of the radio data of 
$\algol$ in the two bands. In the bottom panel (8.3-GHz) the highest 
peak in the frequency interval 0.1--0.9 day$^{-1}$ (only the range 
0.1--0.8 day$^{-1}$ is plotted for presentation reasons) is at 
0.3292$\pm$0.0013 day$^{-1}$. The corresponding periodicity, 
3.037$\pm$0.013 d, is about 6 percent longer than the binary orbital 
period (2.8673 d). The orbital frequency of $\algol$, marked by an 
arrow in the graph does not rise above the noise level. The 
0.33-day$^{-1}$ peak is also seen in the 2.25-GHz band (Fig.~1, top panel) 
at a power similar to a blend of peaks around the orbital period. The 
strong peak at lower frequencies is the $\sim$50-d period ($\sim$0.02 
day$^{-1}$) discovered by \citet{RWG2003}. There is also evidence for 
a signal about twice larger than the 0.33-day$^{-1}$ peak and for the 
beat between the 1-day$^{-1}$ alias and the 0.33-day$^{-1}$ peak. It 
is noted that the power spectrum of the 2.25-GHz band (Fig.~1, top
panel) is very similar to that of the 8.3-GHz data (Fig.~1, bottom
panel), but somewhat noisier. This is a simple consequence of the fact 
that the radio emission in the 8.3-GHz band is stronger. In addition, 
\citet{RWG2003} argued that the 8.3-GHz data represent the mean flux 
more reliably than the 2.25-GHz band.

\begin{figure*}
\epsscale{0.8} 
\plotone{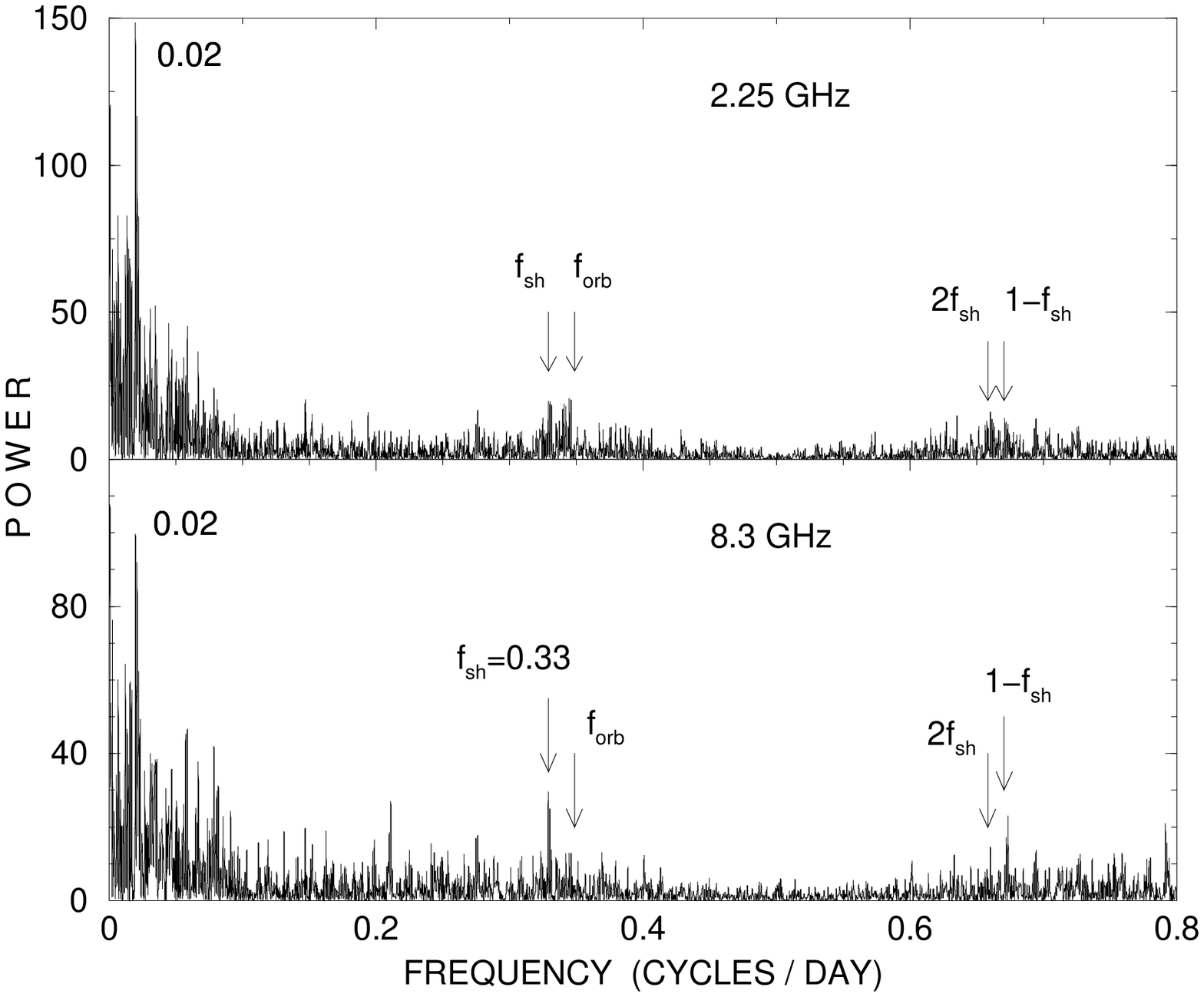}
\caption{Power spectra of the radio data of $\algol$. Top panel: 
2.25-GHz band. Bottom panel: 8.3-GHz data. Note that the radio 
emission in the 8.3-GHz band is stronger than the 2.25-GHz data and 
that the 8.3-GHz data better represent the mean flux \citep{RWG2003}.
Both plots show a low-frequency at about 0.02 day$^{-1}$. In the 
8.3-GHz band, the highest peak in the interval 0.1--0.8 day$^{-1}$ is 
at the frequency f$_{sh}$=0.3292 day$^{-1}$ (which corresponds to a 
period of 3.037$\pm$0.013 d). The location of the orbital frequency 
(which is marked by f$_{orb}$) is also shown. The arrows in the 
right-hand side of the diagram mark the second harmonic of the 
0.3292-day$^{-1}$ peak and its beat frequency with the 1-day$^{-1}$ 
alias.}
\end{figure*}


The 3.037-d candidate periodicity does not appear in the window
function. The significance of this peak in the 8.3-GHz data was 
checked by several methods. First, it was assumed that there are 
no periodic signals in the light curve. The datapoints were randomly 
shuffled and power spectra for 100 simulated light curves were 
computed. The highest peak in these synthetic power spectra in the 
frequency interval 0.1--0.9 day$^{-1}$ reached a power of 11, much 
lower than the value of $\sim$30 of the 0.33-day$^{-1}$ peak in the 
raw data. The frequency range 0--0.1 day$^{-1}$ was rejected because 
the noise sharply rises towards lower frequencies and similarly the 
frequency range near the strong 1-day$^{-1}$ alias (0.9--1.0 day$^{-1}$) 
was avoided. Then, the significance of the suspected periodicity was 
checked in the presence of the 0.02-day$^{-1}$ peak. A sinusoidal 
variation at this frequency was fitted and subtracted from the data. 
The residuals, assumed white noise, were shuffled and added to the 
sinusoid at the 0.02-day$^{-1}$ frequency with the same amplitude 
it has in the data. In 100 simulations, the largest peak in the 
frequency interval 0.1--0.9 day$^{-1}$ reached a power of 12.6, still 
well below the height of the proposed periodicity. In another test, 
it was assumed that the data are modulated with the orbital period 
as well. The significance of the 0.33-day$^{-1}$ peak was similarly 
checked in the presence of this and the 0.02-day$^{-1}$ frequencies. 
Again, the highest peaks in the frequency interval 0.1--0.9 day$^{-1}$ 
in each power spectrum of the 100 simulations were well below the 
observed value. The strongest peak had a power of 13.0 with a standard 
deviation of 1.5, suggesting a very high significance level for the 
candidate periodicity.


In a different test, we estimated the probability of getting a signal 
inside the frequency range adequate for superhumps, which would be 
the strongest peak in the interval 0.1--0.9 day$^{-1}$. The mass ratio 
of $\algol$ implies that a positive superhump (if it exists) should be 
$\sim$4--8\% larger than the orbital period (see previous section) and 
a negative superhump, $\sim$2--4\% shorter \citep{P1999, RHA2003}. This 
implies that that the periodicity found in the light curve of $\algol$ 
is significant at a level of about 97.4\% (1-$f_{orb}\times (\delta 
\epsilon _{+} + \delta \epsilon _{-})/ \delta 
f$=$1-0.34876\times(0.04+0.02)/(0.9-0.1) \approx$0.974, 
where $f_{orb}$ is the orbital frequency, $\delta \epsilon _{+}$ and 
$\delta \epsilon _{-}$ are the permitted ranges of the positive and 
negative superhumps respectively and $\delta f$ is the frequency 
range we examined). The actual significance level is even higher if we 
consider the fact that a similar periodicity was not found in the 
light curves of the other three systems we investigated.


As a final check, the data were divided into two equal sections. The 
first part contained the early points and the second had the later 
datapoints. The candidate periodicity appeared in the power spectra 
of both parts (Fig.~2).

\begin{figure*}
\epsscale{0.8} 
\plotone{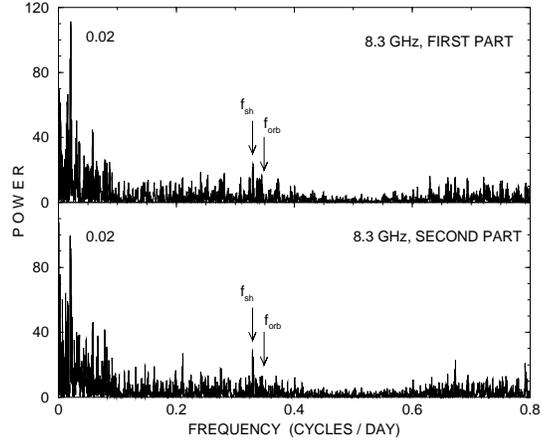}
\caption{Power spectra of two equal sections of the 8.3-GHz data of 
$\algol$. Top panel: first 3721 datapoints. Bottom panel: second part
(last 3721 measurements). The candidate periodicity (f$_{sh}\approx$0.33 
day$^{-1}$) appears in both sections and is the strongest peak in the 
interval 0.1--0.8 day$^{-1}$.}
\end{figure*}

In the power spectrum, the second strongest peak after the 
0.33-day$^{-1}$ frequency is at $\sim$0.21 day$^{-1}$ but it is only
present in the 8.3-GHz data (Fig.~1, bottom panel). Since the noise 
level sharply rises towards low frequencies, it is much less significant 
than the 0.33-day$^{-1}$ frequency. In addition, the fact that it does 
not appear in the two wavelength bands (Fig.~1) nor in both parts of the 
data (Fig.~2) suggests that it is not significant and should be ignored.


We also compared the quiescent data with the flaring observations. 
The radio data of $\algol$ in the 8.3-GHz band were divided into two 
sections. The first part contained the faintest half of the points 
(representing the `quiescent' data) and the second -- the brightest 
measurements (the `flares'). The amplitude cutoff was about 20 mJy. 
The power spectrum of the flares clearly shows the 3-d period, which 
is absent from the power spectrum of the quiescent data (Fig.~3). 

\begin{figure*}
\epsscale{0.8} 
\plotone{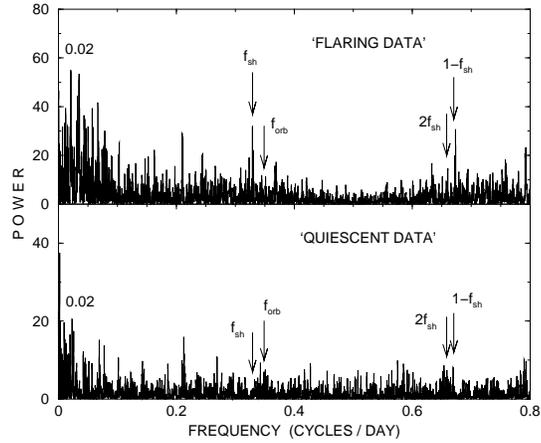}
\caption{Power spectra of the 8.3-GHz observations. 
Top panel: the 3721 brightest datapoints (representing the `flaring' 
data). Bottom panel: the 3721 measurements below about 20 mJy (the 
`quiescent' data). The 0.33 day$^{-1}$ period (f$_{sh}$) is seen in 
the flaring data, but not in quiescence.}
\end{figure*}

Figure 4 presents the mean shape of the 3.037-d period in the `flaring' 
part of the 8.3-GHz data. The peak-to-peak amplitude of the variation 
is about 0.04 Jy.

\begin{figure*}
\epsscale{0.8} 
\plotone{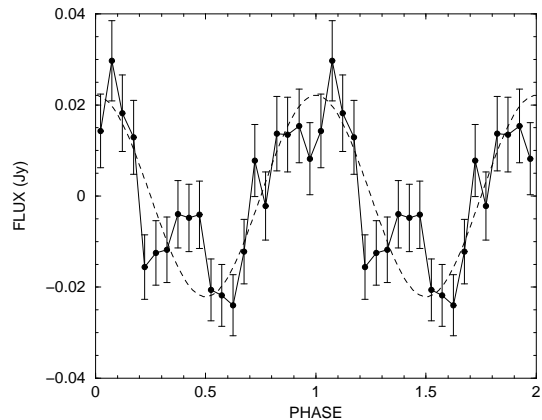}
\caption{The 8.3-GHz brightest datapoints (representing the `flares' 
data) folded on the 3.037-d periodicity and binned into 20 equal 
intervals. Two cycles are shown for clarity. The dashed curve displays 
the sinusoidal fit to the data.}
\end{figure*}

We have also analyzed visual photometry of $\algol$ accumulated by 
the VSNET (Variable Star Network). These data contain 2517 points 
obtained during the past 95 years. No evidence for the 0.33-day$^{-1}$ 
period was found. This result is not surprising since the B star 
overwhelms the light of the binary at these wavelengths.


\section{Discussion}

In this work, radio data from four interacting binary systems were 
re-analyzed. We found evidence for superhumps in the light curve of 
$\algol$ whose mass ratio obeys the theoretical condition for superhumps. 
In the data of the other three systems, we could not find any significant
signals at the expected values (neither near the frequencies calculated
from the theory using the mass ratios nor around the beat periods between
the binary orbital periods and the long-term periodicities). This is as
expected because the RS~CVn systems (V711~Tau and UX~Ari) are detached
binaries (and thus are not expected to possess accretion disks at all)
and since the mass ratio of $\delta$~Lib is slightly above the permitted 
values (Section 1). It is noted that the highest peak around the orbital 
period in the power spectra of V711~Tau and UX~Ari is the orbital period 
itself \citep{RWG2003}, very different than in $\algol$. The power 
spectrum of $\delta$~Lib is very noisy as its coverage is shorter than 
the other three objects and since its radio emission is much weaker than 
the other systems \citep{RWG2003}. Therefore, superhumps could be present
in its data, but with an amplitude below the detection limit.


The power spectrum of $\algol$ shows a periodic signal at 3.037 d,
about 6\% longer than the 2.867-d orbital period. This periodicity 
is located exactly at the beat period between the 50-d period (0.02 
day$^{-1}$) and the orbital period. The tests given in the previous 
section show that it is unlikely caused by chance. Thus, we conclude 
that the phenomenon is real


One may argue that the presence of the short orbital period and a
long-term period may result in some physical change in luminosity
that would be responsible to the presence of the beat periodicity 
in the power spectrum. However, there is only a weak evidence for 
the presence of the orbital period in the radio data of $\algol$ 
(Figs.~1--3). In addition, in the power spectra of the other three 
objects studied in this work there are no significant signals at 
the beat periodicities between their orbital and the long-term 
periods \citep{RWG2003}. These facts indicate that the radio light 
curve of $\algol$ is somewhat different than those of the other 
three systems.

If $\algol$ has superhumps they should be about 4--8\% longer than the 
orbital period (Section 1). The 3.037-d period is at the middle of this 
range. The interpretation of the 3.037-d period as a superhump period 
is tempting. This would mean that the 50-d periodicity represents the 
apsidal precession period of the eccentric accretion disk. We note 
that an extended accretion disk is not required for the formation of 
superhumps. Superhumps have also been found in intermediate polars. 
In this subclass of CVs it is believed that the moderate magnetic 
field of the white dwarf truncates the inner part of the disk 
\citep[e.g.,][and references therein]{RHA2003}. In the following, we 
adopt this interpretation of the data.

\subsection{The mechanism of the radio superhump and the magnetic
activity in Algol systems}

The cause of the observed superhump variation in the optical and X-ray 
light curves of interacting binaries is still under intensive debate. 
\citet{V1982} proposed that the superhump modulation is generated at the 
`bright spot', where the gas stream from the donor star hits the accretion 
disk. \citet{HMB1992} investigated a bright outburst (superoutburst) of 
the dwarf nova OY~Car. They suggested that the superhump variation is 
formed by a large modulation in the surface density of the outer disk that 
results in a change in the form of the bright spot and / or in the 
amount of spillover from the accretion stream. \citet{HHN1992} analyzed 
extensive IUE and Exosat observations of the dwarf nova Z~Cha during 
two superoutbursts and found unexpected dips in its ultraviolet light 
curve. \citet{BMH1996} used HST data during a superoutburst of OY~Car 
and detected similar dips in its ultraviolet data as well. \citet{HHN1992} 
and \citet{BMH1996} thus proposed that the superhump variation is due
to obscuration of the hottest regions of the accretion disk by the
vertical structure at the edge of the disk.

\citet{HKM2001} pointed out that the dissipation of energy in the
accretion disk, which is believed to be the mechanism that gives rise 
to the optical superhump phenomenon observed in CVs, cannot be applied 
to LMXBs. They suggested instead that the disk area is changing with 
the superhump period, and thus predicted that superhumps would mainly 
appear in low inclination systems. \citet{RCB2002} found superhumps in 
X-ray data of the dipping LMXB, V1405 Aql (X1916-053), and offered an 
explanation for the formation of superhumps in high inclination LMXBs. 
According to their model, the thickening of the disk rim, which causes 
an increased obscuration of the X-ray source, is the cause of the 
superhump variation. \citet{RCB2002} further stated that the superhumps 
in LMXBs could be generated by a combination of both scenarios.

The radio emission in Algol consists of a quiescent component and a 
flaring part. Imaging observations indicate that the radio emission 
originates from the polar regions of the cool secondary star. While 
the quiescent emission occupies an extended region above the polar 
zone, flares occur in a more compact region near the stellar surface 
\citep{MMW1998, FMR2000}. The emission is circularly polarized, with 
opposite polarity for the emission from the two polar regions, 
suggesting a dipolar magnetic field structure, and that the emission 
is an optically thin gyrosynchrotron radiation from mildly relativistic 
electrons \citep[for details, see][]{MMW1998}.


The mass-gainer primary in Algol is a late B star whose envelope is 
radiative. The radio emission in Algol systems has thus been ascribed 
to the cool subgiant companion star, which, in contrast, has a thick 
convective envelope where dynamo processes could operate efficiently 
to produce magnetic activity \citep{FMR2000}. At a first glance, it 
is puzzling that the observed 3-d periodicity in the radio light curve 
of $\beta$ Per, if it is a superhump phenomenon as those occur in 
optical and X-ray bands in CVs, is caused by an accretion disk or an 
annulus around the mass-receiver star.

\begin{figure*}
\epsscale{0.8} 
\plotone{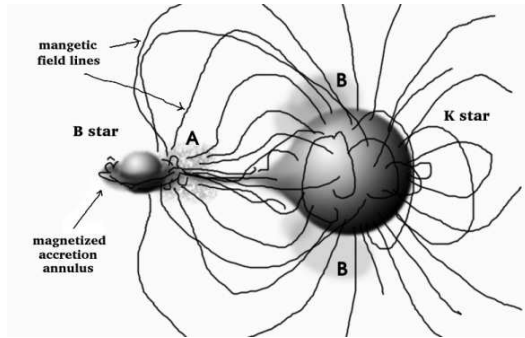}
\caption{An illustration of our proposed schematic magnetic model for
$\algol$ (not in scale). The K star in the system has a strong magnetic
activity. The material that it transfers to the B star is threaded by
magnetic field lines, and the magnetic field carried in the accreting
material is dragged, twisted and amplified in the accretion disk / annulus
around the B star. Reconnection occurs at region A, accelerating electrons
to relativistic energies. These high-energy electrons follow the field
lines, and the dipolar field focuses the electron towards the magnetic
poles (regions B) of the K star, where the gyrosynchrotron radio emission
has the highest intensity.}
\end{figure*}

Here, we propose a scenario to explain the radio properties of Algol 
systems with an accretion disk, and in particular, for the short 
period system $\beta$ Per. It should be noted that for the Algol 
systems with orbital periods shorter than about 5 days, the disk may 
be transient, and unlike accretion disks in compact binaries, the 
velocities within this circumstellar structure around the primary 
star could become sub-Keplerian. This form of accretion was termed 
then `an annulus' (see Section~1). Numerical simulations confirm that 
the mass transfer in the Algol binaries can result in the formation 
of a permanent or transient accretion disk around the primary star 
\citep{BRM1995, RR1998}. This accretion ring/annulus is dense 
and the flow is subsonic. Recent observations show evidence of such a
ring/annulus \citep{RA1999, VHH2001}, supporting the results of the 
numerical simulations and the interpretation of the optical data.




We argue that the accretion disk / annulus is magnetically active. Given
that the secondary star has a significant magnetic field, the material
transferred to the primary star is threaded by magnetic field lines. 
As the ram pressure of the accretion flow is larger than the magnetic 
stress at the Lagrangian point, the ionized accreting material drags 
the field along. Through the accretion disk / annulus, the field lines 
are wound around the mass-receiver star. The field is thus stretched 
and amplified, and at some stage reconnection can occur, causing flaring 
and acceleration of particles. These relativistic particles are trapped 
by the magnetic field of the secondary star and are focused onto the polar 
region by the dipole field. The radio emission is thus greatly enhanced 
in the polar region and becomes detectable in the radio image of the 
system. In spite of the field reconnection in the accretion disk / 
annulus, the lack of radio emission near the B star could be explained by 
the Razin effect and / or plasma low-frequency cut off when the electron 
density is high \citep[see e.g.,][]{D1985}. Figure 5 displays a schematic 
diagram of our model.

\citet{RWG2003} suggested that the frequency of the radio flaring
activity may be correlated with the frequence of changes in the
accretion structure in Algol systems. The superhump period seen in the
flare component of the radio emission can now be explained in terms of
the precession of an asymmetric accretion disk / annulus, which causes
enhanced flare activity when the elongated side of the disk / annulus is
the closest to the secondary star. However, we do not expect to see any
superhump activity in the power spectrum of the quiescent component 
of the emission, consistent with our findings (Section 2, Fig.~3). Our 
model should reproduce a periodicity at about twice the superhump period 
as well. It is interesting to note that such a signal is indeed seen in
the radio light curve of $\algol$ (Figs.~1--3), but this weak feature 
may be alternatively explained by a possible deviation of the mean shape
of the 3.037-d variation from a pure sinusoid (Fig.~4).

Our scenario for the magnetic activity of $\algol$ is analogous to 
that in the RS~CVn systems, where the radio emission is caused by 
the stellar magnetospheric interaction. In fact, the radio properties
of the Algol and the RS~CVn systems are very similar 
\citep[e.g.,][]{RA1993, UTC1998}. In particular, the circular polarization 
in the radio of $\algol$ and the RS~CVn system, V711~Tau (HR 1099), are 
almost identical \citep{MMW1998}. We believe that the only difference 
is that the magnetism in $\algol$ and Algol systems is caused by the
interaction of a magnetic star and a disk / annulus created by the 
field threaded material transferred from the magnetic star, while 
RS~CVn systems like V711~Tau consists of two detached stars both of
which exhibit magnetic activity.

  
\citet{UTC1998} detected radio emission in six out of 26 Algol systems 
surveyed. The radio emission of $\beta$~Per, which probably represents
the Algol systems, is highly variable with strong flares \citep{RWG2003} 
and therefore, the significance of this work is mainly in the statistics. 
We believe that the major reason why so few Algols are not observed with 
strong radio activity is that they are too far away. Other systems may 
be older than expected (e.g., $\delta$~Lib) and older cool stars have 
weaker magnetic fields than young stars of similar spectral type. 
$\beta$~Per is far more active than the other systems simply because 
it is the closest of the Algols, and the strength of the radio flux 
decreases with the square of the distance. We note that the fraction 
of Algols with radio detection is significantly larger than the 
fraction of magnetic accreting systems such as magnetic CVs, among 
which only AE~Aqr and AM~Her have very firm radio emission detections 
\citep{DBC1983, BL1987, BDC1988, BBB1994}. A comparison between 
RS~CVn systems, Algols and magnetic CVs leads us to speculate that 
the flux density could also be related to the size of the regions 
where reconnection occurs and relativistic electrons are accelerated. 
Further observations and theoretical studies will test this hypothesis. 

To summarize, the radio emission in the model that we propose for 
the Algol-like binaries has the following characteristics. Firstly, 
the emission is expected to be strongest in the polar region of the 
secondary star where the field is the largest and the converging 
dipole fields focus the trapped relativistic electrons, which have 
been accelerated by field reconnection in the magnetic disk / annulus 
around the B star. This is in contrast to emission from star spots 
in isolated magnetic stars, which have no strong location preference, 
and to emission from electrons trapped in an equatorial dead zone of 
single magneto-active stars. Secondly, the emission from the two polar 
regions should have different handness (left or right) in the circular 
polarization, indicating the dominant influence of the secondary's
global dipole field, which focuses the relativistic electrons to 
the secondary's poles and interacts with the tangled fields in the 
primary's magnetic disk / annulus. Thirdly, the quiescent component 
of the radio emission should not show the superhump period, as the 
electrons responsible for the quiescent emission are aged electrons 
which have been accumulated in many reconnection and flaring events. 
However, the superhump period could be detected in strong radio 
flares caused by episodic enhancement of field reconnection when the 
asymmetric disk precesses, as there is an injection of a fresh 
population of relativistic electrons.




\subsection{Further considerations}

Initially, the X-ray emission from the Algol systems was connected 
with the accretion between the binary stars \citep{SDE1976, HFT1977}. 
It is now accepted that the origin of the X-ray radiation is the 
secondary cool star \citep[e.g.,][]{WHB1980,FMR2000}. 

Recently, \citet{CDK2004} 
confirmed this concept by finding Doppler shifts caused by the orbital 
motion of the secondary star in several emission lines in Chandra X-ray 
data of $\algol$. The measured shifts were, however, slightly smaller 
than the expected values. Therefore, \citet{CDK2004} suggested that 
either the primary star contributes about 10-15\% of the X-ray light, 
thus shifting the center of mass away from the secondary star or that 
its surrounding corona is distorted towards the primary. The second 
option is consistent with our model for the magnetic activity in 
$\algol$ presented above since the locations of the two B regions at 
the magnetic poles of the K-star in Figure~5 are inclined towards the 
B-type primary star. In line with the ideas presented in our work, we 
suggest that an alternative (or complimentary) simple solution to the 
above inconsistency is that the accretion ring around the primary star 
contributes some of the extra X-ray light that is required in order 
to decrease the Doppler shifts in the emission lines. This proposed 
solution is in agreement with the hydrodynamic results of 
\citet{BRM1995} who found that the disk temperature in Algol 
should get as high as $10^6$ K corresponding to X-ray energy.




Assuming that the observed peak at $\sim$3-d is a true superhump, 
the mass ratio of the binary system of $\algol$ can be estimated 
from the superhump period excess over the orbital period (about 
6\% -- Sections 1--2). From \citet{O1985} we find 
q=M$_{donor}$/M$_{receiver}$=0.29$\pm$0.01. Using the simulations 
presented by \citet{M2000}, a mass ratio of q=0.16$\pm$0.04 is concluded. 
The unweighted mean of these values ($\sim$0.23) is in agreement with 
the previous estimates of q=0.217$\pm$0.005 \citep{HBH1971, TL1978} and 
0.22$\pm0.03$ \citep{RMB1988}. We note, however, that unlike compact 
binaries, in Algol systems the bright extended B star may have some 
gravitational influence on the accretion disk and thus can complicate 
the calculations for superhumps. 

\section{Summary and Conclusions}

We found evidence for a new periodicity about 6\% longer than the orbital 
period in extensive existing radio photometry of $\algol$. The period is 
seen in the flaring data but not in quiescence. This peak is interpreted as 
a positive superhump period. The detection of superhumps in the radio 
light curve of $\algol$ implies that it had some sort of an asymmetrical 
accretion structure that maintained its eccentricity for nearly six 
years. We propose that it is an accretion disk or a ring. This 
suggestion is consistent with interpretation of optical data, Doppler 
tomography and numerical simulations that imply that $\algol$ has an 
accretion annulus around its primary B star. Therefore, Algol systems 
with short orbital periods may have accretion disks / rings / annuli 
that may be more stable than what was previously believed. In our model, 
the 50-d period in the radio data is explained as the apsidal precession 
of the accretion disk / annulus.

Our results extend the superhump phenomenon, which has been observed
in Keplerian accretion disks in compact binaries, to accretion 
structures with sub-Keplerian velocities. It is also the first 
detection of superhumps in an Algol system and the first time
this phenomenon is seen in the radio. 

A new model is proposed for the magnetic activity of Algol-like systems. 
According to our scenario, in addition to the strong magnetic field of 
the secondary star, the magnetic field carried in the accreting material 
is dragged, twisted and amplified in the accretion disk / annulus around 
the primary star. The reconnection of the field lines accelerates 
electrons to relativistic energies and trapping occurs near the magnetic 
poles of the cool star. The binary system is thus reminiscent of RS~CVn 
systems, which consist of two magnetically active stars. This model 
explains several peculiarities in the radio emission of $\algol$. In 
fact, it may further explain the low flaring activity of $\delta$~Lib
\citep{RWG2003} by the fact that it has a binary mass ratio above the 
theoretical limit for superhumps (Sections 1--3).

Our model for the formation of the superhump signal in the radio light 
curves of Algol systems, namely that the reconnection of the field 
lines varies with the phase of the eccentricity, yields a prediction 
that Algol binaries with intermediate orbital periods (thus having a
transient disk or an accretion annulus) and mass ratios below the 
critical value for superhumps, 0.33 (Section 1), should be somewhat 
stronger and harder radio sources than the other objects. Similarly, 
the radiation in other wavebands may be enhanced compared with systems 
with larger mass ratios.

Our scenario for $\algol$ can further explain the fact that the X-ray
Doppler shifts in emission lines of its secondary star are slightly 
smaller than what would be predicted from its orbital rotation. The 
accretion disk annulus is proposed to contribute about 10-15\% of the
total X-ray light. Therefore, we may predict that the X-ray light 
curve of Algol will be modulated with the superhump period as well.

The binary mass ratios in non-eclipsing Algol systems might be 
estimated using the superhump phenomenon. Our results should also 
help to explain some of the observed peculiarities of these systems 
in the infrared, radio, optical, ultraviolet and X-ray bands and 
shed new light on the magnetic activity in binary systems and its 
interaction with the accretion structures.

We hope that future radio and X-ray observations of $\algol$ and other 
Algol-type systems with low mass ratios that should have superhumps 
would be able to confirm or refute our suggestions.

\acknowledgments
This work was partially supported by a postdoctoral fellowship from 
Penn State University. We thank Ariel Marom and Amir Levinson for 
useful discussions and the anonymous referee for his comments. In 
this research, we have used, and acknowledge with thanks, data from 
the VSNET (Variable Star Network) International Databases, based on 
observations submitted by variable star observers worldwide.


\end{document}